\documentclass{Interspeech2024}
\usepackage{multirow}
\usepackage{amssymb} 

\interspeechcameraready

\title{Improved Remixing Process for Domain Adaptation-Based Speech Enhancement by Mitigating Data Imbalance in Signal-to-Noise Ratio}

\name[affiliation={1}]{Li}{Li}
\name[affiliation={1}]{Shogo}{Seki}

\address{
  $^1$CyberAgent, Inc., Japan
\email{\{li\_li, seki\_shogo\}@cyberagent.co.jp}}


\def\Mat#1{\boldsymbol{\mathbf{#1}}}

\def\Vec#1{\textsf{\boldmath $#1$}}

\def\E{{\mathbb E}}

\def\R{{\mathbb R}}
\def\T{{\textsf{T}}}

\def\thline{\noalign{\hrule height 1.2pt}}
\DeclareMathSymbol{\shortminus}{\mathbin}{AMSa}{"39}

\newcommand{\reffig}[1]{Fig. \ref{fig:#1}}

\newcommand{\reftab}[1]{Table \ref{tab:#1}}

\begin{document}

\maketitle

\begin{abstract}
    
    RemixIT and Remixed2Remixed are domain adaptation-based speech enhancement (DASE) methods that use a teacher model trained in full supervision to generate pseudo-paired data by remixing the outputs of the teacher model. The student model for enhancing real-world recorded signals is trained using the pseudo-paired data without ground truth. Since the noisy signals are recorded in natural environments, the dataset inevitably suffers data imbalance in some acoustic properties, leading to subpar performance for the underrepresented data. The signal-to-noise ratio (SNR), inherently balanced in supervised learning, is a prime example. In this paper, we provide empirical evidence that the SNR of pseudo data has a significant impact on model performance using the dataset of the CHiME-7 UDASE task, highlighting the importance of balanced SNR in DASE. Furthermore, we propose adopting curriculum learning to encompass a broad range of SNRs to boost performance for underrepresented data.
    
\end{abstract}

\section{Introduction}
Speech enhancement (SE) \cite{loizou2007speech} is a technique that improves the quality of recorded speech in the presence of noise and interference, having a wide range of practical applications. Recent advances in deep neural networks (DNNs) have significantly boosted the capabilities of SE systems \cite{ochieng2023deep}. Particularly, SE models trained in full supervision \cite{luo2019conv,hu2020dccrn,tzinis2020sudo,yu2022dbt} have achieved impressive performance in numerous benchmarks. However, when faced with real-world recorded signals, these models suffer from performance degradation due to a distribution mismatch between the synthetic training data and recorded data.
\par
Several methods have recently been proposed to tackle this issue, which can be categorized into two primary concepts: methods that use accessible signal characteristics and metrics instead of clean speech to guide model training from scratch and the utilization of domain adaptation methods to transition the domain of the training data (source domain) to the recorded data (target domain). 
The former category includes approaches such as the utilization of positive-unlabeled learning (PLUSE) \cite{ito2023audio}, the replacement of clean target with noisy target as the ground truth (NyTT) \cite{fujimura2021noisy}, and the training of models by optimizing evaluation metrics \cite{subramanian2019speech,fu2022metricgan} or observation consistency \cite{wisdom2020unsupervised,saijo2023self}.
The latter category includes approaches that leverage teacher-student learning (a.k.a., knowledge distillation) \cite{knowledge} to generate pseudo-paired data using the teacher model, which is employed to train the student model.
To acquire high-performance models with less data, we follow approaches in the latter category, domain adaptation-based speech enhancement (DASE). 
\par
RemixIT \cite{tzinis2022remixit} and Remixed2Remixed (Re2Re) \cite{li2023remixed2remixed} are two recently proposed DASE methods. 
Specifically, RemixIT and Re2Re 
utilize supervised learning models trained on synthetic noisy-clean pair speech as the teacher model. 
The student model initialized with the teacher model is then updated using pseudo-paired data generated by remixing the speech and noise signals estimated by the teacher model. The key differences between the two methods exist in the composition of the generated pair data and the loss function used for training the student model.
RemixIT applies the remixing process once to generate pseudo-noisy-clean pair data and uses a reconstruction loss between the signals predicted by the teacher and student models.
Conversely, Re2Re applies the remixing process twice to generate pseudo-noisy-noisy pair data and employs the Noise2Noise learning \cite{lehtinen2018noise2noise}. 
Regardless of the differences, 
both methods have demonstrated superior performance in DASE tasks.
\par
In this paper, we improve RemixIT and Re2Re by concentrating on the remixing process, a pivotal element in both methods. 
In conventional methods, the remixing is conducted without any manual intervention, which could lead to an imbalanced dataset for the student model training, resulting in a suboptimal model. 
One crucial characteristic is the signal-to-noise ratio (SNR), which is inherently balanced as a standard process when synthesizing data in supervised learning but overlooked in the DASE task. 
Generally, there are two primary strategies to learn from such imbalanced datasets \cite{krawczyk2016learning,branco2016survey}: the data pre-processing approaches and special-purpose learning methods. 
Considering that the distribution of SNR can be easily adjusted during the remixing process, we opt for the data pre-processing approach, which aims to alter the data distribution so that standard training algorithms can be adopted.
To manage this aspect effectively, we introduce an SNR control module (SNRCM) into the remixing process. 
Furthermore, we propose adopting curriculum learning (CL) \cite{Wang2021A} to cover a broad range of SNRs since the preliminary experiments revealed difficulties associated with training models across a broad range of SNRs.

\section{Remixing-based DASE}
\label{sec:DASE}
\subsection{Common training strategy}
RemixIT and Re2Re employ a teacher-student learning strategy, which consists of a teacher model $\mathcal{F}_{\mathcal{T}}(\theta_\mathcal{T})$ and a student model $\mathcal{F}_{\mathcal{S}}(\theta_\mathcal{S})$. Here, $\theta_\mathcal{T}$ and $\theta_\mathcal{S}$ are parameters of the teacher and student models, respectively. The teacher model is trained in supervision using synthetic noisy-clean pair data $(\Vec{x}, \Vec{s}, \Vec{n})$ by minimizing the reconstruction error of speech and noise signals,
where $\Vec{s}, \Vec{n}, \Vec{x}=\Vec{s}+\Vec{n}$ denote clean speech, noise, and noisy speech signals, respectively. 
The student model is first initialized with parameters of the pre-trained teacher model and then further trained to enhance the real-world recorded data with only the recorded noisy data $\Vec{x}'\sim \mathcal{D}_{\Vec{x}'}$ accessible.
Given a mini-batch of noisy data $\Mat{x}'=\Mat{s}'+\Mat{n}'\in\R^{B\times T}$, the teacher model estimates the speech and noise signals as follows:
\begin{align}
\tilde{\Mat{s}}', \tilde{\Mat{n}}' = \mathcal{F}_{\mathcal{T}}(\Mat{x}'; \theta_{\mathcal{T}}^{(k)}),
\end{align}
where $\cdot^{(k)}$ denotes the $k$-th epoch and the bold Roman font represents a batch $\Mat{a}=[\Vec{a}_1, \ldots, \Vec{a}_B]^\T$ including multiple signals $\Vec{a}_b$ drawn from distribution $\mathcal{D}_{\Vec{a}}$. Here, $^\T$ denotes the transpose operator, and $B$ and $T$ denote the mini-batch size and signal length, respectively. 
The estimated noise signals are then shuffled and remixed with the estimated speech signals to generate the pseudo-paired data for updating $\theta_\mathcal{S}^{(k)}$. The teacher model is continuously updated during the training phase using the weighted moving average (WMA), expressed as $\theta_{\mathcal{T}}^{(k+1)} = \gamma \theta_{\mathcal{S}}^{(k)} + (1-\gamma) \theta_{\mathcal{T}}^{(k)}$, to generate more accurate pseudo-paired data. 
Here, $0\leq\gamma\leq 1$ is the weight parameter.

\subsection{RemixIT}
RemixIT generates pseudo-noisy-clean pair data $(\tilde{\Mat{x}}', \tilde{\Mat{s}}', \tilde{\Mat{n}}')$, where the
bootstrapped mixture $\tilde{\Mat{x}}'$ is obtained by remixing
$\tilde{\Mat{x}}'=\tilde{\Mat{s}}' + \Mat{P}\tilde{\Mat{n}}'$.
Here, $\Mat{P}\sim \Pi_{B\times B}$ is a permutation matrix to shuffle the estimated noise signals in each batch.
The student model $\mathcal{F}_{\mathcal{S}}$ is trained by minimizing the reconstructed error between the outputs of the model and the pseudo-targets $\tilde{\Mat{s}}'$ and $\tilde{\Mat{n}}'$ as follows:
\begin{align}
\hat{\Mat{s}}', \hat{\Mat{n}}' &= \mathcal{F}_{\mathcal{S}}(\tilde{\Mat{x}}'; \theta_{\mathcal{S}}^{(k)}), \\
\mathcal{L}_{\rm RemixIT} &= \sum_{b=1}^B \big[
\mathcal{L}(\hat{\Mat{s}}'_b, \tilde{\Mat{s}}'_b) + \mathcal{L}(\hat{\Mat{n}}'_b, [\Mat{P}\tilde{\Mat{n}}']_b) \big].
\label{eq:RemixIT}
\end{align}

\subsection{Remixed2Remixed}
Re2Re generates pseudo-noisy-noisy pair data $(\bar{\Mat{x}}', \tilde{\Mat{x}}')$, where $\tilde{\Mat{x}}'$ is the same as the one in the RemixIT
and $\bar{\Mat{x}}'$ is given by 
$\bar{\Mat{x}}' = \tilde{\Mat{s}}' + \Mat{Q}\tilde{\Mat{n}}'$.
Here, $\Mat{Q}$ is another permutation matrix following $\Mat{Q}\sim \Pi_{B\times B}$.
The student model is trained using the Noise2Noise learning \cite{lehtinen2018noise2noise}, whose loss function is given by
\begin{align}
\mathcal{L}_{\rm Re2Re} =  
\E_{(\bar{\Mat{x}}',\tilde{\Mat{x}}')}
\big[\mathcal{L}(\hat{\Mat{s}}', \bar{\Mat{x}}')\big].
\label{eq:N2N}
\end{align}

\section{Imbalanced dataset analysis}
\label{sec:proposed}
In this section, we first present empirical evidence to raise the issue that datasets for student model training generated via the remixing process are imbalanced with a skewed SNR distribution. Although this analysis is performed on the CHiME-7 UDASE task dataset \cite{leglaive2023chime} as a representative example, real-world recorded datasets without manual modification inevitably face such data imbalance.  Following this, we introduce an SNR control module (SNRCM) and curriculum learning (CL) \cite{Wang2021A}. These strategies are designed to enhance the model performance across a vast range of SNRs, particularly for data underrepresented within the skewed distribution.

\subsection{Brief introduction to the UDASE training dataset}
The UDASE task comprises three datasets: (1) the LibriMix paired dataset for training supervised SE model and development; (2) the CHiME-5 unlabeled recorded dataset for adopting domain adaptation, development, and evaluation; and (3) the reverberant LibriCHiME-5 close-to-in-domain paired dataset for development and evaluation. 
Here, we focus on the CHiME-5 unlabeled dataset mainly utilized for domain adaptation training. 
The CHiME-5 dataset \cite{barker2018fifth} was recorded at 4-person dinner parties, which comprised noisy multi-speaker utterances of 20 English conversation sessions. 
The CHiME-7 UDASE task excerpted the utterances where participants wearing microphones did not speak (i.e., the maximum number of simultaneously active speakers was three) and divided the 20 sessions for training ($\approx$83h), development ($\approx$15.5h), and evaluation ($\approx$7h), respectively. 
Training data was segmented into chunks up to 10s, and a pre-trained voice activity detector (VAD) was used for data pre-processing. This resulted in two versions of the training dataset: CHiME-5 w/o VAD and CHiME-5 w/ VAD.

\subsection{SNR distributions of original and remixed datasets}
\begin{figure}[t]
    \centering
    \includegraphics[width=\linewidth]{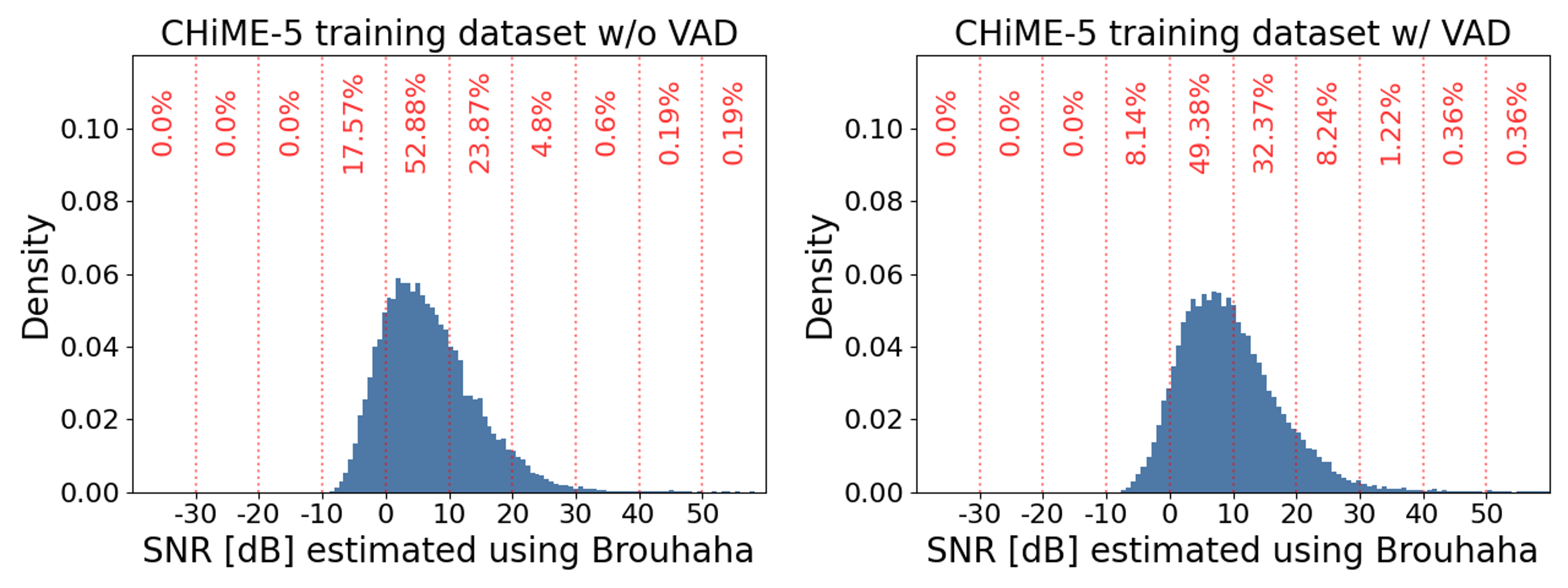}
    \caption{Estimated SNR distributions for CHiME-5 training dataset w/o VAD (left) and w/ VAD (right).}
    \label{fig:SNR_dist}
    \vspace{-6pt}
\end{figure}
\begin{figure}[t]
    \centering
    \includegraphics[width=\linewidth]{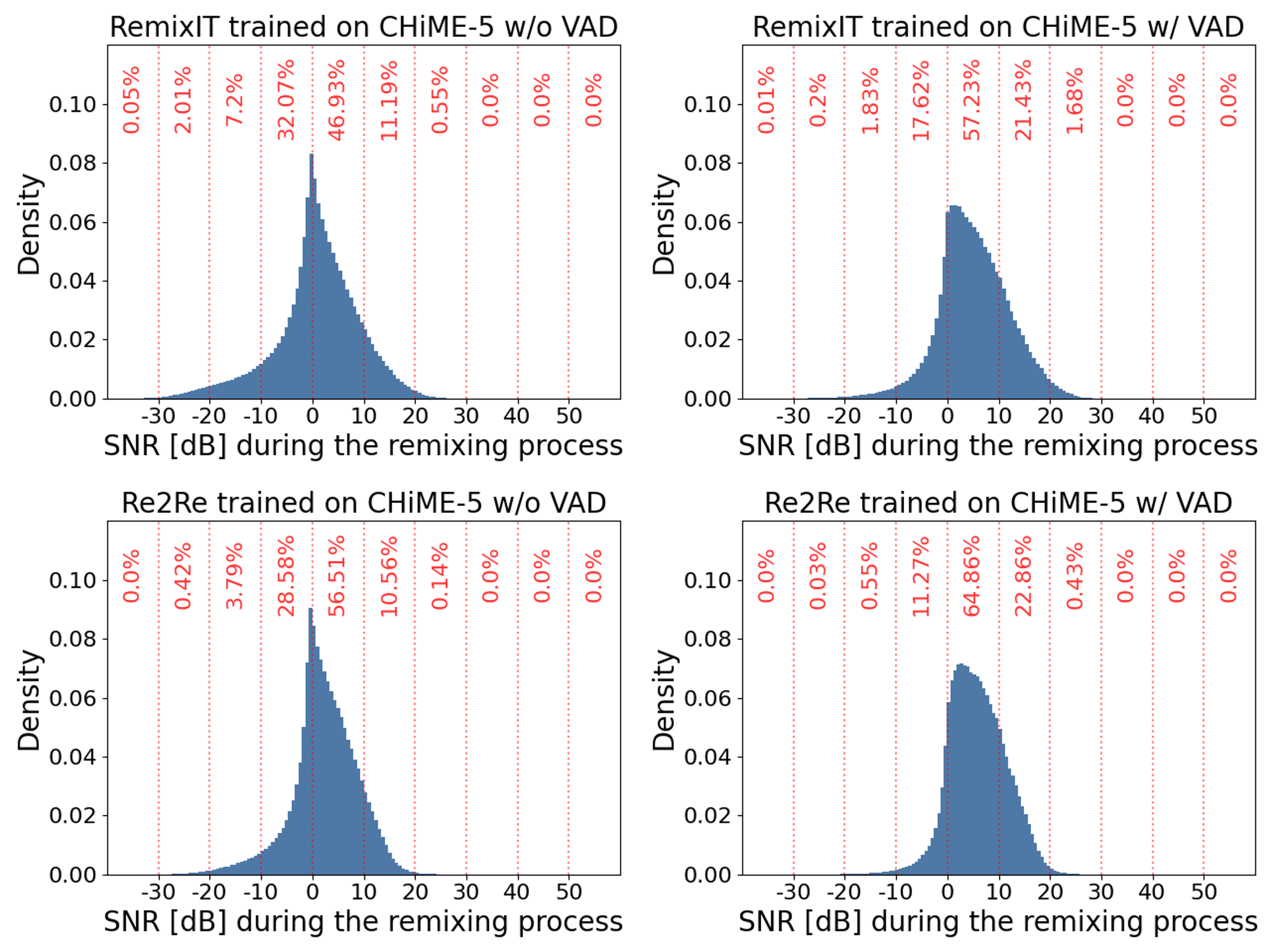}
    \caption{Measured SNR distributions for datasets generated by the remixing process in RemixIT (1st row) and Re2Re (2nd row), respectively. The left and right columns correspond to models trained on CHiME-5 w/o VAD and w/ VAD, respectively.}
    \label{fig:SNR_dist2}
    \vspace{-6pt}
\end{figure}

To obtain the SNR distributions for the aforementioned training datasets, we utilized Brouhaha \cite{lavechin2023brouhaha}, a multi-task model for VAD, SNR, and C50 (a measure of speech clarity) room acoustics estimation, as in the CHiME-7 UDASE task. Brouhaha is trained using approximately 1,250 hours of synthetic signals generated by contaminating clean speech segments with silence, noise, and reverberation.
Note that Brouhaha is trained using segments with a single speaker, while the CHiME-5 dataset contains segments with up to three active speakers.
The estimated SNR distributions for the CHiME-5 training datasets are depicted in \reffig{SNR_dist}. 
Both distributions are right-skewed, 
peaking within the range of (0, 10].
The second most data-rich range is (10, 20], which constitutes roughly 76.75\% and 81.75\% of the entire datasets when combined with the most data-rich range.
The remaining roughly 20\% of the data spans the broad ranges of (-10, 0] and (20, 60].
Compared to the range of (0, 20], these data are significantly underrepresented in the overall dataset. 
This leads the trained model to tend to optimize for data within (0, 20] and may be suboptimal for these underrepresented data.
However, the underrepresented data also appear in inference, as the test data in domain adaptation tasks is assumed to align closely with the training data.
\par
For RemixIT and Re2Re, we measured SNRs for all remixed noisy signals. The results of these measurements are illustrated in \reffig{SNR_dist2}.
Similarly, these distributions are skewed. 
Using VAD as pre-processing increased the quantity of data in the range of (0, 20], specifically from 58\% to 77\% for RemixIT and from 67\% to 88\% for Re2Re, bringing it closer to the original training dataset. This distribution shift increases the amount of data within (0, 20] that the student model is exposed to, which could lead to improved performance for the data within this SNR range. As a result, scores may be improved when evaluated on a dataset with a similar SNR distribution.

\subsection{SNR-aware remixing}
\begin{figure}
    \centering
    \includegraphics[width=\linewidth]{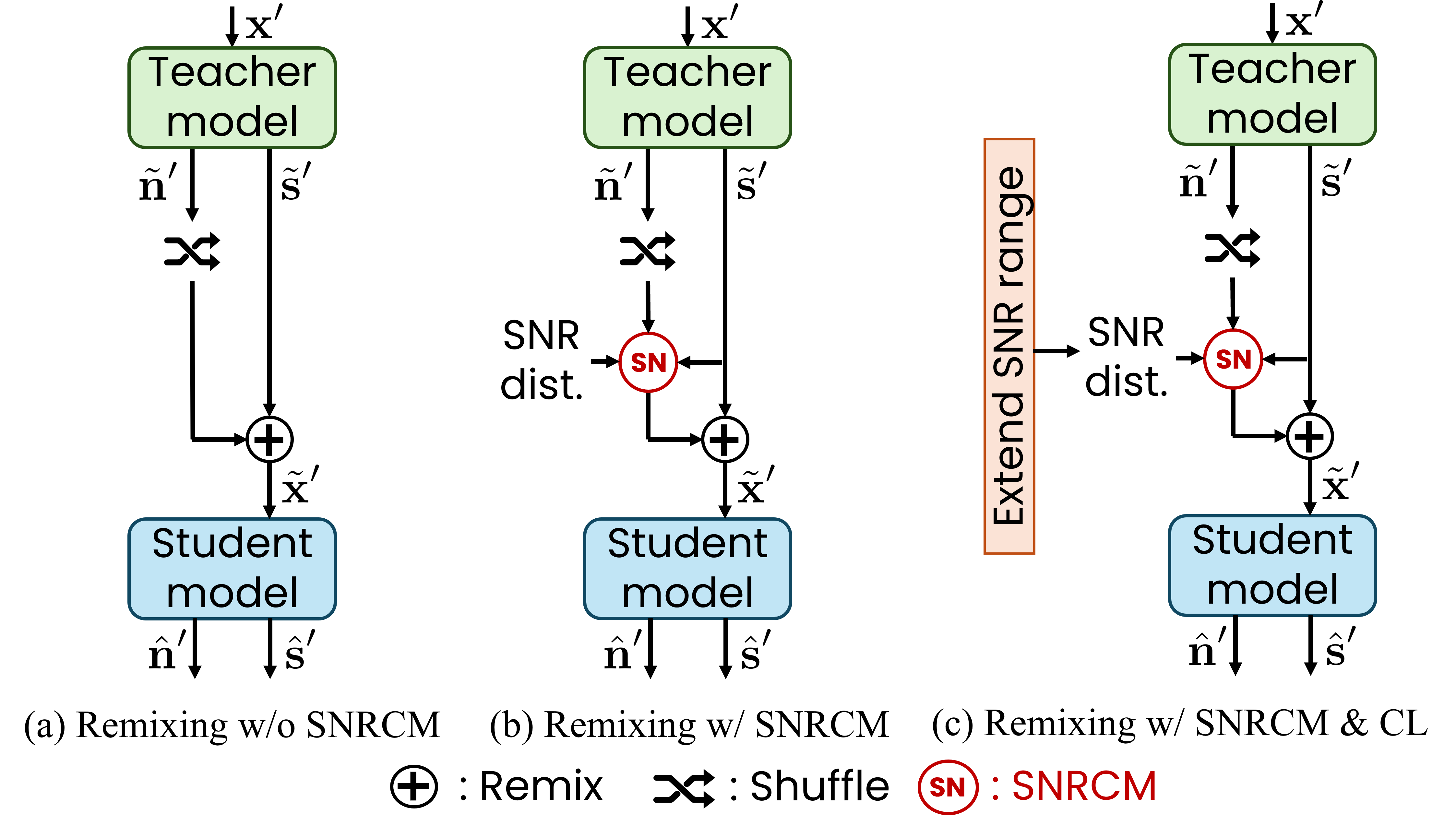}
    \caption{Flowcharts of (a) remixing without SNRCM, (b) remixing with SNRCM using predefined SNR distribution, and (c) remixing with SNRCM and CL that extends the range of SNR distribution in each training stage.}
    \label{fig:methods}
    \vspace{-7pt}
\end{figure}
To improve such suboptimal models caused by imbalanced datasets, we incorporate an SNRCM into the remixing processes of RemixIT and Re2Re. This module randomly samples an SNR from a predefined balanced distribution and then remixes the noise and speech signals to meet the sampled SNR. There are various ways to define this distribution. Here, we opt for a uniform distribution as it can be applied to all datasets without tuning.
The uniform distribution boundaries are highly dependent on the dataset and practical applications.
The wider the range of SNRs, the more difficult it is to train the model. There is a trade-off between the generalization ability of the model with respect to SNRs and the difficulty of model training.
For applications where the objective is to optimize performance for frequently occurring data, and performance for infrequently occurring data is less important, it is advisable to select an SNR range that covers most data while keeping the SNR range relatively narrow, e.g., 20 to 30 dB.
Conversely, choosing an SNR range that covers the entire dataset is crucial for applications that require a decent level of performance for all data. However, in this case, it becomes important to increase training data or optimize the training methods to achieve good generalizability.
Since the issue of poor generalization capability has been observed in our preliminary experiments for SNR range spanning 40 dB to 50 dB, we propose adopting CL \cite{Wang2021A} to increase generalization capability.
In particular, we divide the entire training phase into several stages and use multiple SNR ranges. The SNR range for the initial stage is set to the most data-rich range of approximately 20 to 30 dB and gradually increases at each stage. 
Note that the SNR distribution throughout the entire training phase using CL is no longer uniform. \reffig{methods} illustrates the flowchart of remixing in conventional methods and those in the proposed methods. 

\section{Experimental evaluations}
\label{sec:experiment}
\begin{figure*}
    \centering
    \includegraphics[width=0.98\linewidth]{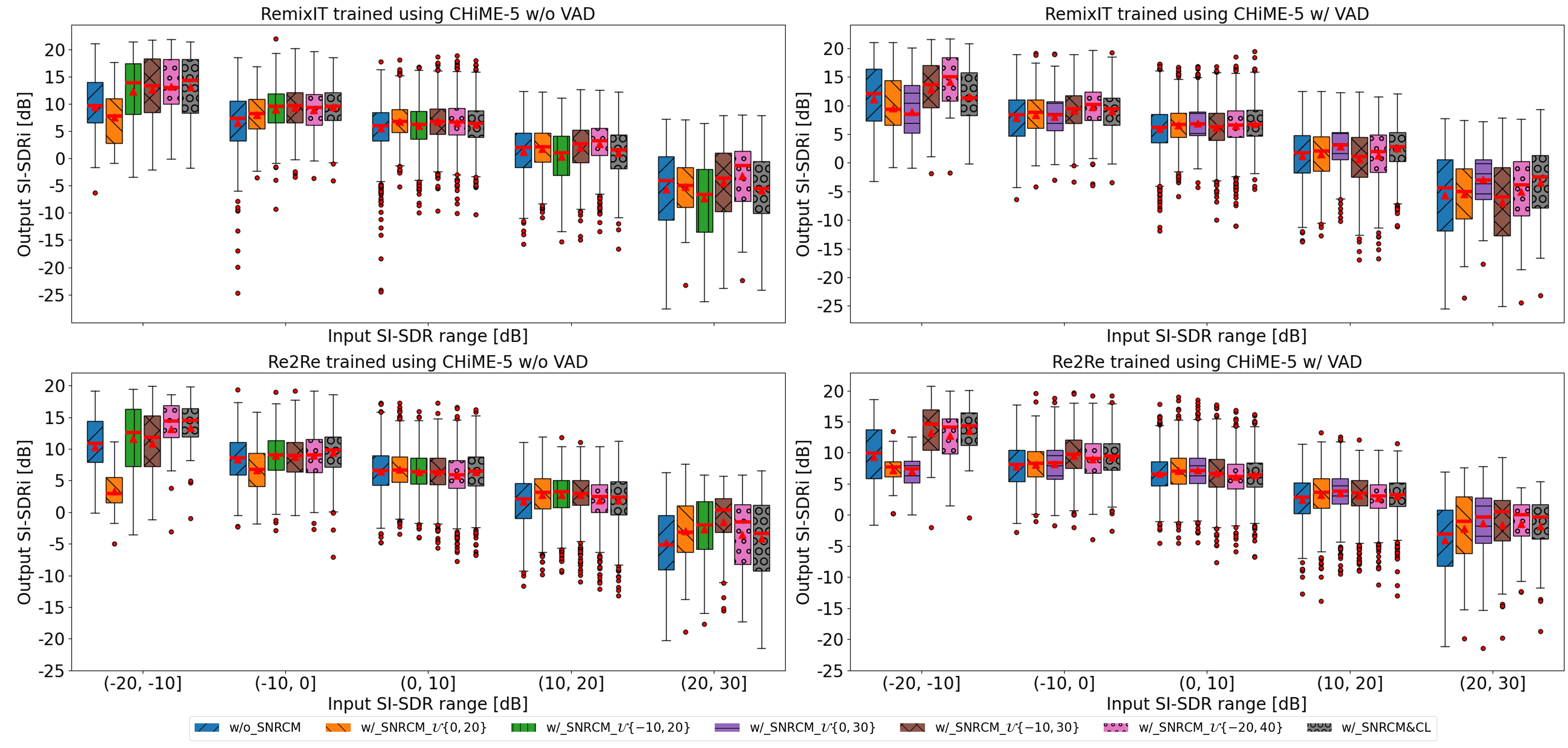}
    \caption{SI-SDR improvement [dB] achieved by RemixIT (top) and Re2Re (bottom). Models were trained with CHiME-5 w/o VAD (left) and w/ VAD (right), respectively. The red lines represent the median values, and the red triangle marks indicate the mean values. Teacher and student models were initialized using the checkpoint provided by the CHiME-7 UDASE task.}
    \label{fig:boxplot}
    \vspace{-6pt}
\end{figure*}

\subsection{Evaluation dataset and  metrics}
In this subsection, we provide more information about the reverberant LibriCHiME-5 close-to-in-domain dataset used for evaluation. 
This dataset is a synthetic dataset of reverberant noisy speech labeled with clean speech. The clean speech and noise signals were excerpted from the LibriSpeech \cite{panayotov2015librispeech} and the noise-only segments in the CHiME-5 dataset, respectively.
Room impulse responses (RIRs) were excerpted from the VoiceHome corpus \cite{voicehome}, recorded in the living room, kitchen, and bedroom of three real homes.
The mixtures were generated by adding noise segments to randomly sampled speech utterances convolved with randomly sampled RIRs. The SNR for each speaker was distributed as a Gaussian distribution $\mathcal{N}(5, 7)$ to match the original CHiME-5 dataset. The proportions of the subsets labeled with the maximum number of active speakers
were 0.6, 0.35, and 0.05, respectively.
The data durations for evaluation were approximately 3 hours, including 1952 samples. 
We used three objective scores, scale-invariant signal-to-distortion ratio (SI-SDR) \cite{le2019sdr}, the perceptual evaluation of speech quality (PESQ) \cite{rix2001perceptual,pesq}, and the short-time intelligibility index (STOI) \cite{taal2010short,stoi}, as the evaluation metrics according to 
the analysis of the relationship between objective and subjective evaluation metrics conducted by the organizers of the CHiME-7 UDASE task \cite{leglaive2024objective}. They found that nonintrusive metrics, such as DNSMOS \cite{dnsmos} and TorchAudio-Squim \cite{torchaudio} measured with the CHiME-5 test dataset, demonstrated less correlation than intrusive metrics computed on the LibriCHiME-5 dataset. As a result, we opted only to evaluate the LibriCHiME-5 dataset.

\subsection{Model architecture and training settings}
We followed the baseline training script provided by the CHiME-7 UDASE task \cite{leglaive2023chime}.
The Sudo rm-rf \cite{tzinis2020sudo} architecture was used for both the teacher and student models. 
The encoder and decoder of these models consisted of one-dimensional convolution and transpose convolution, respectively, with 512 filters of 41 taps and a hop size of 20 samples. The separator was composed of 8 U-Conv blocks. The pre-trained teacher model was used to initialize the student model and was continually updated by WMA with a weight of $\gamma=0.01$ every epoch. The batch size and the number of training epochs were set at 24 and 200, respectively.
The negative SI-SDR \cite{le2019sdr} and mean squared error (MSE) was employed as the loss function for training the student model in RemixIT and Re2Re, respectively.
Based on the analyzed SNR distributions of the CHiME-5 training dataset, we selected the uniform distributions with ranges of 20, 30, and 40 dB, which contain the most data in each set, as the predefined SNR distribution. Namely, $\mathcal{U}\{0, 20\}$, $\mathcal{U}\{\shortminus10, 20\}$, and $\mathcal{U}\{\shortminus10, 30\}$ for the CHiME-5 w/o VAD, and $\mathcal{U}\{0, 20\}$, $\mathcal{U}\{0, 30\}$, and $\mathcal{U}\{\shortminus10, 30\}$ for the CHiME-5 w/ VAD.
We selected a uniform distribution with an extended range of 10dB on both sides as the most comprehensive range for evaluation, i.e., $\mathcal{U}\{\shortminus20, 40\}$.
For CL, we divided the 200 epochs into four stages, allocating 50 epochs to each stage. 
The SNR range for the initial stage spanned 30 dB, specifically $\mathcal{U}\{\shortminus 10, 20\}$ for CHiME-5 w/o VAD and $\mathcal{U}\{0, 30\}$ for CHiME-5 w/ VAD. As the training stage progressed, the SNR range gradually increased to $\mathcal{U}\{\shortminus10, 30\}$, $\mathcal{U}\{\shortminus10, 40\}$, and finally $\mathcal{U}\{\shortminus15, 45\}$, reaching the most comprehensive range of 60 dB.

\subsection{Experimental results and discussions}
\begin{table}[t]
\centering
\caption{Objective evaluation scores in LibriCHiME-5 test dataset averaged over 4 student model initializations. ``conv." and ``prop." denote conventional methods without SNRCM and proposed methods using SNRCM and CL. The presence of ``VAD" indicates the version of the CHiME-5 training dataset. }
\label{tab:result}
\setlength{\tabcolsep}{3.5pt}
\begin{tabular}{@{\extracolsep{1pt}}lcccccc}
\thline
        & \multicolumn{2}{c}{SI-SDR {[}dB{]}} & \multicolumn{2}{c}{PESQ} & \multicolumn{2}{c}{STOI} \\
\cline{2-3}\cline{4-5}\cline{6-7}
\multicolumn{1}{l}{\multirow{-2}{*}{Methods}}        & conv.       & prop.       & conv.  & prop. & conv.  & prop. \\
\hline
RemixIT         & 11.66 & \bf{12.19} & 1.83 & \bf{1.85} & 0.82  & \bf{0.83} \\
RemixIT-VAD     & 11.81 & \bf{12.50} & 1.79 & \bf{1.84} & 0.81  & \bf{0.83} \\
Re2Re           & 12.01 & \bf{12.13} & \bf{1.86} & 1.84 & 0.82  & 0.82      \\
Re2Re-VAD       & 12.22 & \bf{12.57} & \bf{1.88} & 1.87 & 0.82  & \bf{0.83} \\
\hline
Input  \cite{leglaive2024objective}        &  \multicolumn{2}{c}{6.60}  & \multicolumn{2}{c}{1.55} & \multicolumn{2}{c}{0.71} \\
N\&B \cite{leglaive2024objective}       &  \multicolumn{2}{c}{13.00}  & \multicolumn{2}{c}{2.40} & \multicolumn{2}{c}{0.80} \\
\thline
\end{tabular}
\vspace{-7pt}
\end{table}

\reffig{boxplot} illustrates the SI-SDR improvement [dB] achieved by RemixIT and Re2Re trained with CHiME-5 w/o VAD and w/ VAD, respectively.
The results show that the narrow SNR range of 20-30 dB improved model performance for data within that specific range, especially when CHiME-5 w/ VAD was used for training data, but significantly degraded performance for data outside that range. 
For the entire evaluation dataset, which contains approximately 78\% of the data in the (0, 20], the RemixIT and Re2Re models trained using SNRCM with $\mathcal{U}\{0, 20\}$ on the dataset without VAD achieved SI-SDR improvements of 1.03 dB and 0.08 dB respectively, compared to models not using SNRCM. Meanwhile, models trained on the dataset with VAD achieved SI-SDR improvements of 0.47 dB and 0.55 dB, respectively.
In a broader SNR range of 40 dB or more (three boxes from the right), models trained with SNRCM consistently achieved better or comparable performance than models without SNRCM. When applied to different methods and training datasets, these three settings yielded different trends across various input SNR ranges. However, the method employing CL consistently achieved moderate performance on average.
These results indicate that the SNR of the remixed noisy speech significantly impacts model performance, and the SNRCM effectively increases the controllability of model performance.
\par
\reftab{result} summarizes the objective metrics averaged over four student model initializations, including the checkpoint provided by the CHiME-7 UDASE task and three teacher models trained from scratch with random seeds. 
The results showed that the proposed method with SNRCM and CL significantly improved SI-SDR for RemixIT and slightly for Re2Re. However, there was no noticeable improvement in PESQ and STOI. 
One reason for the smaller improvement in Re2Re compared to RemixIT is that the remixed noisy signal is already more evenly distributed in the (0,20] range. 
This reduces the effectiveness of SNRCM, finally leading to comparable scores between the two methods.
Compared to the top-ranked system in the challenge (N\&B), the differences in SI-SDR were reduced to 0.5 dB and 0.43 dB for RemixIT and Re2Re, respectively.
While STOI was slightly higher, PESQ was lower.
We consider the task of determining the reason for the discrepancy between PESQ and the other two metrics as future work.

\section{Conclusions}
\label{sec:conclusion}
This paper highlighted the issue of imbalanced datasets in remixing-based DASE models and demonstrated the adverse impact of skewed SNR distributions using the CHiME-7 UDASE task dataset. We balanced the dataset by integrating an SNR control module and increased model generalization by employing curriculum learning. We validated the effectiveness of the proposed method through experimental evaluations.

\newpage
\bibliographystyle{IEEEtran}
\bibliography{mybib}

\begin{thebibliography}{10}
\providecommand{\url}[1]{#1}
\csname url@samestyle\endcsname
\providecommand{\newblock}{\relax}
\providecommand{\bibinfo}[2]{#2}
\providecommand{\BIBentrySTDinterwordspacing}{\spaceskip=0pt\relax}
\providecommand{\BIBentryALTinterwordstretchfactor}{4}
\providecommand{\BIBentryALTinterwordspacing}{\spaceskip=\fontdimen2\font plus
\BIBentryALTinterwordstretchfactor\fontdimen3\font minus
  \fontdimen4\font\relax}
\providecommand{\BIBforeignlanguage}[2]{{%
\expandafter\ifx\csname l@#1\endcsname\relax
\typeout{** WARNING: IEEEtran.bst: No hyphenation pattern has been}%
\typeout{** loaded for the language `#1'. Using the pattern for}%
\typeout{** the default language instead.}%
\else
\language=\csname l@#1\endcsname
\fi
#2}}
\providecommand{\BIBdecl}{\relax}
\BIBdecl

\bibitem{loizou2007speech}
P.~C. Loizou, \emph{Speech enhancement: Theory and practice}.\hskip 1em plus
  0.5em minus 0.4em\relax CRC press, 2007.

\bibitem{ochieng2023deep}
P.~Ochieng, ``Deep neural network techniques for monaural speech enhancement
  and separation: State of the art analysis,'' \emph{Artificial Intelligence
  Review}, vol.~56, no. Suppl 3, pp. 3651--3703, 2023.

\bibitem{luo2019conv}
Y.~Luo and N.~Mesgarani, ``{Conv-TasNet}: Surpassing ideal time-frequency
  magnitude masking for speech separation,'' \emph{IEEE/ACM transactions on
  audio, speech, and language processing}, vol.~27, no.~8, pp. 1256--1266,
  2019.

\bibitem{hu2020dccrn}
Y.~Hu, Y.~Liu, S.~Lv, M.~Xing, S.~Zhang, Y.~Fu, J.~Wu, B.~Zhang, and L.~Xie,
  ``{DCCRN}: Deep complex convolution recurrent network for phase-aware speech
  enhancement,'' \emph{arXiv preprint arXiv:2008.00264}, 2020.

\bibitem{tzinis2020sudo}
E.~Tzinis, Z.~Wang, and P.~Smaragdis, ``{Sudo RM-RF}: Efficient networks for
  universal audio source separation,'' in \emph{Proc. IEEE International
  Workshop on Machine Learning for Signal Processing (MLSP)}, 2020, pp. 1--6.

\bibitem{yu2022dbt}
G.~Yu, A.~Li, H.~Wang, Y.~Wang, Y.~Ke, and C.~Zheng, ``{DBT-Net}: Dual-branch
  federative magnitude and phase estimation with attention-in-attention
  transformer for monaural speech enhancement,'' \emph{IEEE/ACM Transactions on
  Audio, Speech, and Language Processing}, vol.~30, pp. 2629--2644, 2022.

\bibitem{ito2023audio}
N.~Ito and M.~Sugiyama, ``Audio signal enhancement with learning from positive
  and unlabeled data,'' in \emph{Proc. IEEE International Conference on
  Acoustics, Speech and Signal Processing (ICASSP)}, 2023, pp. 1--5.

\bibitem{fujimura2021noisy}
T.~Fujimura, Y.~Koizumi, K.~Yatabe, and R.~Miyazaki, ``Noisy-target training: A
  training strategy for dnn-based speech enhancement without clean speech,'' in
  \emph{Proc. IEEE European Signal Processing Conference (EUSIPCO)}, 2021, pp.
  436--440.

\bibitem{subramanian2019speech}
A.~S. Subramanian, X.~Wang, M.~K. Baskar, S.~Watanabe, T.~Taniguchi, D.~Tran,
  and Y.~Fujita, ``Speech enhancement using end-to-end speech recognition
  objectives,'' in \emph{Proc. IEEE Workshop on Applications of Signal
  Processing to Audio and Acoustics (WASPAA)}, 2019, pp. 234--238.

\bibitem{fu2022metricgan}
S.-W. Fu, C.~Yu, K.-H. Hung, M.~Ravanelli, and Y.~Tsao, ``{MetricGAN-U}:
  Unsupervised speech enhancement/dereverberation based only on
  noisy/reverberated speech,'' in \emph{Proc. IEEE International Conference on
  Acoustics, Speech and Signal Processing (ICASSP)}, 2022, pp. 7412--7416.

\bibitem{wisdom2020unsupervised}
S.~Wisdom, E.~Tzinis, H.~Erdogan, R.~Weiss, K.~Wilson, and J.~Hershey,
  ``Unsupervised sound separation using mixture invariant training,''
  \emph{Advances in Neural Information Processing Systems}, vol.~33, pp.
  3846--3857, 2020.

\bibitem{saijo2023self}
K.~Saijo and T.~Ogawa, ``{Self-Remixing}: Unsupervised speech separation via
  separation and remixing,'' in \emph{Proc. IEEE International Conference on
  Acoustics, Speech and Signal Processing (ICASSP)}, 2023, pp. 1--5.

\bibitem{knowledge}
G.~Hinton, O.~Vinyals, and J.~Dean, ``Distilling the knowledge in a neural
  network,'' in \emph{Proc. NIPS Deep Learning and Representation Learning
  Workshop}, 2015.

\bibitem{tzinis2022remixit}
E.~Tzinis, Y.~Adi, V.~K. Ithapu, B.~Xu, P.~Smaragdis, and A.~Kumar,
  ``{RemixIT}: Continual self-training of speech enhancement models via
  bootstrapped remixing,'' \emph{IEEE Journal of Selected Topics in Signal
  Processing}, vol.~16, no.~6, pp. 1329--1341, 2022.

\bibitem{li2023remixed2remixed}
L.~Li and S.~Seki, ``{Remixed2Remixed}: Domain adaptation for speech
  enhancement by {Noise2Noise} learning with remixing,'' in \emph{Proc. IEEE
  International Conference on Acoustics, Speech and Signal Processing
  (ICASSP)}, 2024, pp. 806--810.

\bibitem{lehtinen2018noise2noise}
J.~Lehtinen, J.~Munkberg, J.~Hasselgren, S.~Laine, T.~Karras, M.~Aittala, and
  T.~Aila, ``{Noise2Noise}: Learning image restoration without clean data,'' in
  \emph{Proc. International Conference on Machine Learning (ICML)}, 2018, pp.
  2965--2974.

\bibitem{krawczyk2016learning}
B.~Krawczyk, ``Learning from imbalanced data: Open challenges and future
  directions,'' \emph{Progress in Artificial Intelligence}, vol.~5, no.~4, pp.
  221--232, 2016.

\bibitem{branco2016survey}
P.~Branco, L.~Torgo, and R.~P. Ribeiro, ``A survey of predictive modeling on
  imbalanced domains,'' \emph{ACM computing surveys (CSUR)}, vol.~49, no.~2,
  pp. 1--50, 2016.

\bibitem{Wang2021A}
X.~Wang, Y.~Chen, and W.~Zhu, ``A survey on curriculum learning,'' \emph{IEEE
  Transactions on Pattern Analysis and Machine Intelligence}, vol.~44, pp.
  4555--4576, 2021.

\bibitem{leglaive2023chime}
S.~Leglaive, L.~Borne, E.~Tzinis, M.~Sadeghi, M.~Fraticelli, S.~Wisdom,
  M.~Pariente, D.~Pressnitzer, and J.~R. Hershey, ``The {CHiME-7 UDASE} task:
  Unsupervised domain adaptation for conversational speech enhancement,'' in
  \emph{Proc. 7th International Workshop on Speech Processing in Everyday
  Environments (CHiME)}, 2023.

\bibitem{barker2018fifth}
J.~Barker, S.~Watanabe, E.~Vincent, and J.~Trmal, ``The fifth `{CHiME}' speech
  separation and recognition challenge: Dataset, task and baselines,'' in
  \emph{Proc. Interspeech}, 2018.

\bibitem{lavechin2023brouhaha}
M.~Lavechin, M.~M{\'e}tais, H.~Titeux, A.~Boissonnet, J.~Copet, M.~Rivi{\`e}re,
  E.~Bergelson, A.~Cristia, E.~Dupoux, and H.~Bredin, ``Brouhaha: Multi-task
  training for voice activity detection, speech-to-noise ratio, and {C50} room
  acoustics estimation,'' in \emph{Proc. IEEE Automatic Speech Recognition and
  Understanding (ASRU)}, 2023.

\bibitem{panayotov2015librispeech}
V.~Panayotov, G.~Chen, D.~Povey, and S.~Khudanpur, ``{LibriSpeech}: An {ASR}
  corpus based on public domain audio books,'' in \emph{Proc. IEEE
  international conference on acoustics, speech and signal processing
  (ICASSP)}, 2015, pp. 5206--5210.

\bibitem{voicehome}
\BIBentryALTinterwordspacing
N.~Bertin, E.~Camberlein, R.~Lebarbenchon, S.~Peillon, E.~Lamandé,
  S.~Sivasankaran, F.~Bimbot, I.~Illina, A.~Tom, S.~Fleury, and E.~Jamet,
  ``{VoiceHome} corpus: A corpus dedicated to distant-microphone speech
  processing in domestic environments,'' 2018. [Online]. Available:
  \url{https://doi.org/10.5281/zenodo.1252143}
\BIBentrySTDinterwordspacing

\bibitem{le2019sdr}
J.~Le~Roux, S.~Wisdom, H.~Erdogan, and J.~R. Hershey, ``{SDR} -- half-baked or
  well done?'' in \emph{Proc. IEEE International Conference on Acoustics,
  Speech and Signal Processing (ICASSP)}, 2019, pp. 626--630.

\bibitem{rix2001perceptual}
A.~W. Rix, J.~G. Beerends, M.~P. Hollier, and A.~P. Hekstra, ``Perceptual
  evaluation of speech quality ({PESQ}) -- a new method for speech quality
  assessment of telephone networks and codecs,'' in \emph{Proc. IEEE
  International Conference on Acoustics, Speech, and Signal Processing
  (ICASSP)}, vol.~2, 2001, pp. 749--752.

\bibitem{pesq}
\BIBentryALTinterwordspacing
M.~Wang, C.~Boeddeker, R.~G. Dantas, and ananda seelan, ``{ludlows/python-pesq:
  supporting for multiprocessing features},'' May 2022. [Online]. Available:
  \url{https://doi.org/10.5281/zenodo.6549559}
\BIBentrySTDinterwordspacing

\bibitem{taal2010short}
C.~H. Taal, R.~C. Hendriks, R.~Heusdens, and J.~Jensen, ``A short-time
  objective intelligibility measure for time-frequency weighted noisy speech,''
  in \emph{Proc. IEEE International Conference on Acoustics, Speech and Signal
  Processing (ICASSP)}, 2010, pp. 4214--4217.

\bibitem{stoi}
\BIBentryALTinterwordspacing
M.~Pariente. [Online]. Available: \url{https://github.com/mpariente/pystoi}
\BIBentrySTDinterwordspacing

\bibitem{leglaive2024objective}
S.~Leglaive, M.~Fraticelli, H.~ElGhazaly, L.~Borne, M.~Sadeghi, S.~Wisdom,
  M.~Pariente, J.~R. Hershey, D.~Pressnitzer, and J.~P. Barker, ``Objective and
  subjective evaluation of speech enhancement methods in the {UDASE} task of
  the 7th {CHiME} challenge,'' \emph{arXiv preprint arXiv:2402.01413}, 2024.

\bibitem{dnsmos}
C.~K. Reddy, V.~Gopal, and R.~Cutler, ``{DNSMOS P. 835}: A non-intrusive
  perceptual objective speech quality metric to evaluate noise suppressors,''
  in \emph{Proc. IEEE International Conference on Acoustics, Speech and Signal
  Processing (ICASSP)}, 2022, pp. 886--890.

\bibitem{torchaudio}
A.~Kumar, K.~Tan, Z.~Ni, P.~Manocha, X.~Zhang, E.~Henderson, and B.~Xu,
  ``Torchaudio-squim: Reference-less speech quality and intelligibility
  measures in torchaudio,'' in \emph{Proc. IEEE International Conference on
  Acoustics, Speech and Signal Processing (ICASSP)}, 2023, pp. 1--5.

\end{thebibliography}

\end{document}